# Self-Assembled Ligand-Capped Plasmonic Au Nanoparticle Films in the Kretschmann Configuration for Sensing of Volatile Organic Compounds

Rituraj Borah[1,3†], Jorid Smets[1,4†], Rajeshreddy Ninakanti[1,2,3], Max L. Tietze[4], Rob Ameloot[4], Dmitry N. Chigrin[5,6], Sara Bals[2,3], Silvia Lenaerts[1,3], Sammy W. Verbruggen[1,3*]

[1]Sustainable Energy, Air & Water Technology (DuEL), Department of Bioscience Engineering, University of Antwerp, Groenenborgerlaan 171, 2020, Antwerp, Belgium

[2]Electron Microscopy for Material Science (EMAT) Department of Physics, University of Antwerp, Groenenborgerlaan 171, 2020, Antwerp, Belgium

[3]NANOlab Center of Excellence, University of Antwerp, Groenenborgerlaan 171, 2020, Antwerp, Belgium

[4]Centre for Membrane Separations, Adsorption, Catalysis, and Spectroscopy (cMACS), KU Leuven - University of Leuven, Celestijnenlaan 200F, Leuven, 3001, Belgium

[5]DWI – Leibniz-Institut für Interaktive Materialien e.V., Forckenbeckstraße 50, 52056 Aachen, Germany

[6]Institute of Physics (1A), RWTH Aachen University, Sommerfeldstr. 14, 52074 Aachen, Germany

[†]These authors contributed equally

*Corresponding author: Sammy.Verbruggen@uantwerpen.be

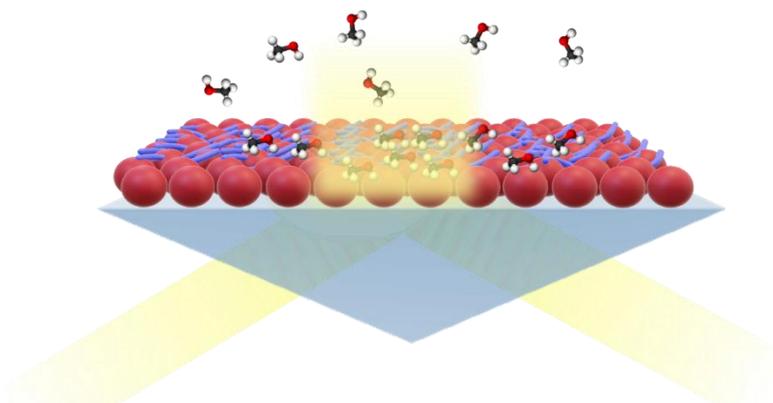

## Abstract

Films of close-packed Au nanoparticles are coupled electrodynamically through their collective plasmon resonances. This collective optical response results in enhanced light-matter interactions, which can be exploited in various applications. Here, we demonstrate their application in sensing volatile organic compounds, using methanol as a test-case. Ordered films over several cm$^2$ were obtained by interfacial self-assembly of colloidal Au nanoparticles (~10 nm diameter) through controlled evaporation of the solvent. Even though isolated nanoparticles of this size are inherently non-scattering, when arranged in a close-packed film the plasmonic coupling results in a strong reflectance and absorbance. The *in-situ* tracking of vapor phase methanol concentration through UV-Vis transmission measurements of the nanoparticle film is first demonstrated. Next, *in-situ* ellipsometry of the self-assembled films in the Kretschmann (also known as ATR) configuration is shown to yield enhanced sensitivity, especially with phase difference measurements, Δ. Our study shows the excellent agreement between theoretical models of the spectral response of self-assembled films with experimental *in-situ* sensing experiments. At the same time, the theoretical framework provides the basis for the interpretation of the various observed experimental trends. Combining periodic nanoparticle films with ellipsometry in the Kretschmann configuration is a promising strategy towards highly sensitive and selective plasmonic thin-film devices based on colloidal fabrication methods for volatile organic compound (VOC) sensing applications.

**Introduction**

In multiple applications of nanoparticles, ordered periodic arrays are particularly interesting since the light-matter interaction is enhanced in such periodic assemblies. Given an optimal spacing between the nanoparticles, this effect occurs due to the coupling of the individual plasmon polaritons, resulting in an amplified coherent optical response.[1][2] To exploit this plasmonic coupling, periodic arrays of nanostructures have gained considerable attention.[3] For instance, plasmonic nanostructure arrays have been demonstrated for light-harvesting enhancement in photovoltaic applications[4], photocatalytic processes[5], photothermal mediated processes[6], etc. Furthermore, the intense near-field enhancement upon plasmonic coupling in closely packed nanostructure arrays or clusters results in a significant improvement in surface-enhanced Raman spectroscopy (SERS)[7][8][9], fluorescence[10], surface-enhanced infrared absorption spectroscopy (SEIRA)[11], photoemission and photodetection[12] signals.

Self-assembly is a suitable technique to obtain periodic arrays of nanoparticles in a simple and cost-efficient way.[13][14] The use of tailored nanoparticles as building blocks allows application-specific functionalities in which the macro-scale interface or bulk phase properties arise from the individual nanoparticles.[15][16][17] In recent years, self-assembled films of nanoparticles have gained attention, both in terms of fabrication techniques and novel applications.[18][19] Air-liquid interfacial assembly, a technique originally devised for monolayer films of organic molecules, is also a promising route towards nanoparticle arrays. Current approaches based on air-water interfacial assembly, such as Langmuir-Blodgett troughs, involve precise control of the interfacial pressure through compression and expansion of the interfacial area to tune of the film morphology. An alternative liquid phase is ethylene glycol, which has been extensively used by, *e.g.*, Murray and co-workers, for obtaining monolayer films as well as binary nanoparticle arrays with nanoparticles of two different sizes.[20][21]

In self-assembly techniques, improved control over parameters such as the packing configuration, inter-particle gap, number of layers, *etc.*, facilitates tailoring towards specific applications.[7][22][23][24][25] In particular, self-assembled plasmonic nanoparticle (*e.g.* Au nanoparticles) films are interesting for sensing applications because of the high sensitivity of the plasmon excitation to the local dielectric environment.[26][27][28] Importantly, to excite the surface plasmon polaritons in planar films, the Kretschmann configuration is commonly used to compensate for the momentum mismatch between the incident photon and the plasmon polariton wave.[29] In this work, self-assembled Au nanoparticle films fabricated at the air-ethylene glycol interface are used as a sensing platform. The films consist of discrete 10 nm oleylamine-capped Au nanoparticles that create a ligand-nanoparticle network-like structure and exhibit collective coupled plasmon resonances due to the strong inter-particle coupling which can be excited by both normal incident electromagnetic waves, and incidence following the Kretschmann configuration. *In-situ* optical transmission measurements using normal incidence and the Kretschmann configurations were compared. In the Kretschmann configuration set-up, *in-situ* ellipsometry was integrated in order to measure the relative optical response of the film to *p*- and *s*-polarized incident light in the presence of the controlled dosing of methanol vapor. The same methanol dosing procedure was also implemented in the *in-situ* transmission spectroscopy set-up. It is shown that our homogeneous ligand-capped plasmonic nanoparticle film obtained through facile wet chemical procedures, holds the potential for a powerful volatile organic compound (VOC) sensing technology, that may be expanded towards other application domains such as *e.g.* biosensing.

**Results and discussion**

**Self-assembly of colloidal Au nanoparticles**

In this work, a gold nanoparticle self-assembly procedure at the air-ethylene glycol interface is extended to fabricate homogeneous films of several cm$^2$, as opposed to previously reported studies reaching only several mm$^2$ films.[20][30]

The advantage of ethylene glycol as a subphase is that, unlike water, there is no need for careful regulation of the surface pressure for the self-assembly of small (~10 nm) nanoparticles. As shown in **Figure 1 (a)**, upon the evaporation of the solvent (toluene), the nanoparticles get trapped at the interface due to non-polarity of the surface ligands that does not allow the nanoparticles to migrate to the bulk liquid. Pohjalainen *et al*. showed that colloidal nanoparticles, especially with excess ligands, spread with much greater ease on ethylene glycol than water, with significantly lower surface pressure.[31] Following the existing literature on large area air-water interfacial assembly with larger nanoparticles (>100 nm) over a fixed water surface,[32][33] the initial experiments in this work with water as a subphase (and small 10 nm Au nanoparticles) failed due to apparent migration of nanoparticles to the walls, and formation of clusters of nanoparticles and oleylamine ligands in patches. Due to the relatively low polarity and surface tension, ethylene glycol slows down this process, thus facilitating gradual re-arrangement of the nanoparticles and ligands at the interface. This allows formation of close-packed arrays, while at the same time preventing the nanoparticles and ligands to migrate (or dissolve) into the bulk. As shown in **Figure 1 (b)**, a glass beaker half-filled with ethylene glycol was sealed leaving a small hole for the slow evaporation of toluene over *ca.* 8 hours. This approach ensures well-controlled convection that induces slow evaporation of toluene from the opposite side, resulting in gradual film formation (**Figure 1 (d)**). **Figure 1 (d)** shows the development of a self-assembled film which is predominantly monolayer, from 20 mL (~$10^{14}$ nanoparticles) of the as-synthesized nanoparticle colloid concentrated into 400 µL in the toluene phase. As shown in the low magnification SEM image in **Figure 1 (c)** and **S6**, the films obtained after transferring the self-assembled layers to a glass substrate by the Langmuir-Blodgett deposition technique, are uniform over large areas. The nanoparticles arrange themselves into hexagonal close packed arrays to form predominantly mono- or multi-layered films depending on the concentration of the nanoparticles in the colloid suspension. Although the as-synthesized nanoparticle colloid is deep red in color, the film appears blue due to the interparticle coupling which red-shifts the collective plasmon resonance. The conversion of the optical spectra of colloidal nanoparticles and the nanoparticle films to simulated color distributions as shown in **Figure 1 (d)** matches the observed colors changes indicating the formation of the film from the colloidal nanoparticles. Excess nanoparticles accumulate at specific locations, leaving most of the film highly uniform. Importantly, excess nanoparticles or ligand in the toluene phase create an extra surface pressure at the interface that helps the formation of the well-ordered film.[34] In the conventional Langmuir-Blodgett self-assembly process, the surface pressure increases as the nanoparticle film is gradually compressed, in a way similar to a gas volume compression.[35] The present procedure, however, has a fixed surface area upon which the nanoparticles are introduced. Thus, as the toluene phase evaporates leaving the nanoparticles trapped at the interface, the surface pressure increases. As the surface pressure is generally low for ethylene glycol, the excess ligands increases the surface pressure and facilitates a more controlled self-assembly process. The reproducibility of the self-assembly procedure was ascertained by repeating the experiment and using containers of different sizes, *i.e.,* varying interfacial area and nanoparticle concentrations (see **Figure 1(e)**).

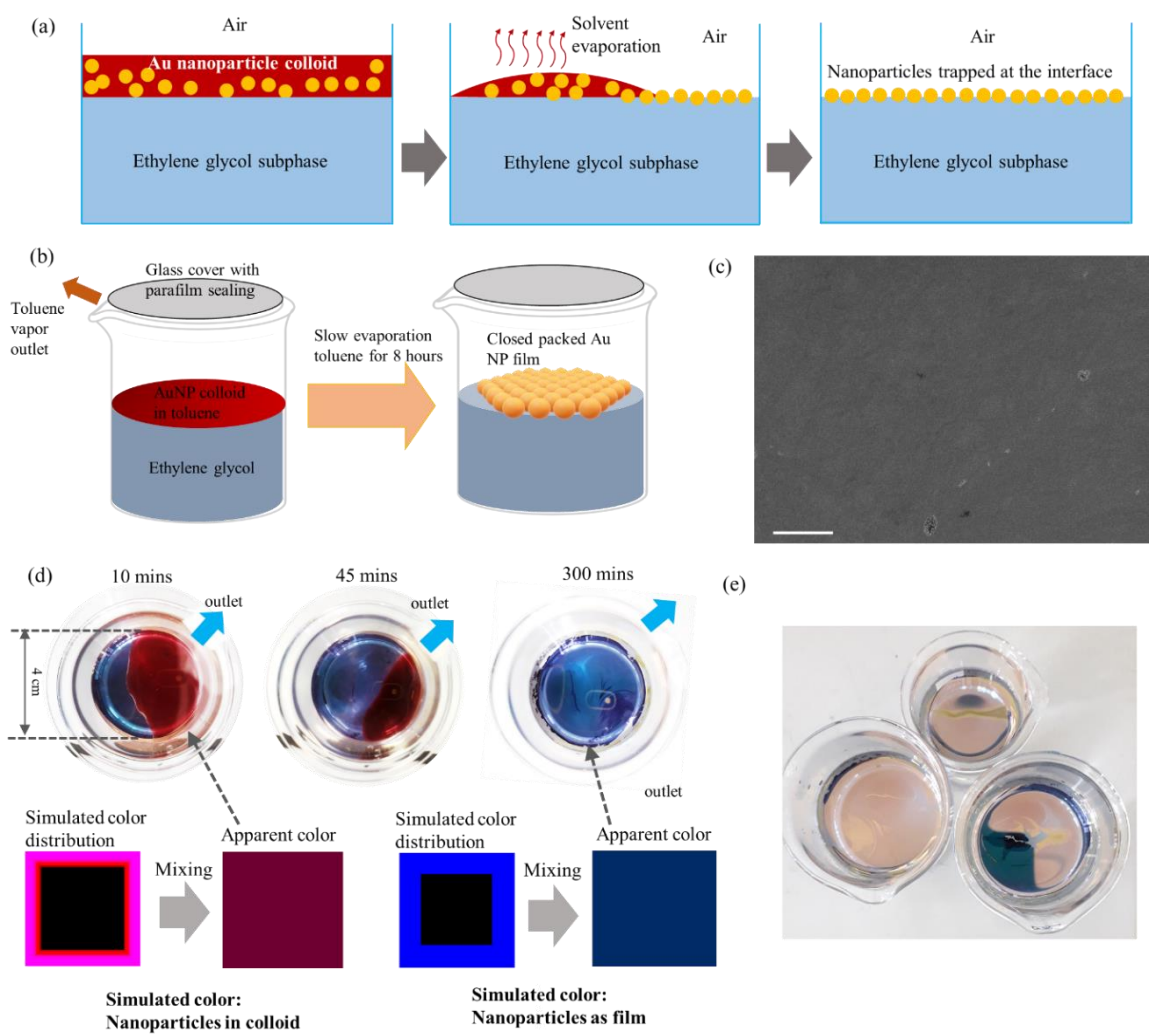

**Figure 1**. (a) Schematic representation of the air-ethylene glycol interfacial self-assembly by trapping of nanoparticles at the interface. (b) Schematic representation the self-assembly set-up for controlled ethylene glycol evaporation. (c) SEM image showing the uniformity of the films over large areas (scale bar: 2 µm). (d) The time evolution of the film-formation as the solvent toluene evaporates. Also, simulated RGB color distribution of colloidal Au nanoparticles and self-assembled Au nanoparticle films obtained by converting the optical spectra to colors by a MATLAB function (supporting information). (e) Picture of beakers after film formation.

**Description of the optical configurations and modeling**

The collective resonance by plasmonic coupling in periodic Au nanostructures or films can be described adequately by classical electrodynamics.[1,36] In order to excite the surface plasmon polaritons in a planar Au thin film, it is necessary to implement specific geometric arrangements such as the Kretschmann configuration (or the Otto configuration) to compensate for the momentum mismatch between the incident photon and the plasmon polariton wave.[37] A nanoparticle film, on the contrary, does not necessarily require such specific geometrical arrangements, *i.e.* there is no stringent need of a denser medium for the incident beam with an incident angle larger than the critical angle. In fact, the collective plasmon resonances in the lattice can be excited at direct normal incidence, where the collective response can arise from near-field coupling, far-field coupling, or diffractive coupling.[1,38,39] In the case of small 10 nm nanoparticles in this work, the individual localized surface plasmon modes (LSPR) of the close packed nanoparticles couple through near-field overlap to result in a collective resonance of the film or lattice. The excitation of the collective plasmon resonances is also strongly determined by the incident angle of light. As shown in **Figure 2 (a)**, as the incident angle $\theta$ varies, the plasmonic coupling and the optical response also vary, depending on the E-field polarization (defined by the angle $\varphi$). For $p$-polarized ($\varphi = 90°$) and $s$-polarized ($\varphi = 0°$) incidence at an interface between two dielectric media, the reflectance and transmittance can be defined by the classical law of Fresnel reflection.[40] However, the optical response for the nanoparticle film results from the excitation of the collective plasmon resonances. The theoretical description of these resonances requires rigorous numerical solution of the Maxwell's equations or theoretical approximation due to the complex structure. The shiny appearance of these films (**Figure 1 (b, c)**) when viewed from an angle, indicates the reflection by a collective optical response. For the electromagnetic modeling of the optical response of the films, a unit cell is constructed as the computational domain with Floquet periodic boundary conditions on the side walls to account for the extension of the film in 2D to infinity (see **Figure 2 (b)**). The schematic in **Figure 2 (c)** shows the Au nanoparticle film in the Kretschmann (or ATR) configuration. In the Kretschmann configuration, a high refractive index is required from the side of incidence, the nanoparticle film is located at the other side where a lower refractive index is required. The incoming beam is incident at an angle higher than the critical reflection angle, to allow total internal reflection and the resulting standing evanescent wave on the lighter medium interacts with the object of interest. Thus, in contrast to normal direct incidence, the collective plasmon resonance of the nanoparticle film in the Kretschmann configuration is excited by the evanescent field. Also, only the light absorbed by the object of interest is absent from the reflected beam and no light is transmitted. It is important to note that the evanescent field excitation from the $p$- and $s$-polarized incident light has completely different directionalities when studying nanoparticle films. While for $p$-polarized incidence, the evanescent field has both in-plane (plane of the film) and out-of-plane components, for $s$-polarized incidence, the evanescent field is confined to the plane of the film.[41] To verify the accuracy of the used electromagnetic models, results from literature were reproduced. The agreement of the results from Mueller *et al.*[42] (finite difference time domain, FDTD) with our FEM models indicates the validity of our approach for the normal incidence case (**Figure S3**). The electromagnetic model of the Kretschmann configuration was first validated by comparing the computed reflectance and transmittance with analytical solutions obtained from Fresnel's coefficients in the absence of the Au nanoparticles (**Figure S4 (a)**). Additionally, the angle-resolved reflectance of an Au thin film in the Kretschmann configuration (experimental data from Vohnsen *et al*.[43]) was satisfactorily reproduced by our model (**Figure S4 (b)**).

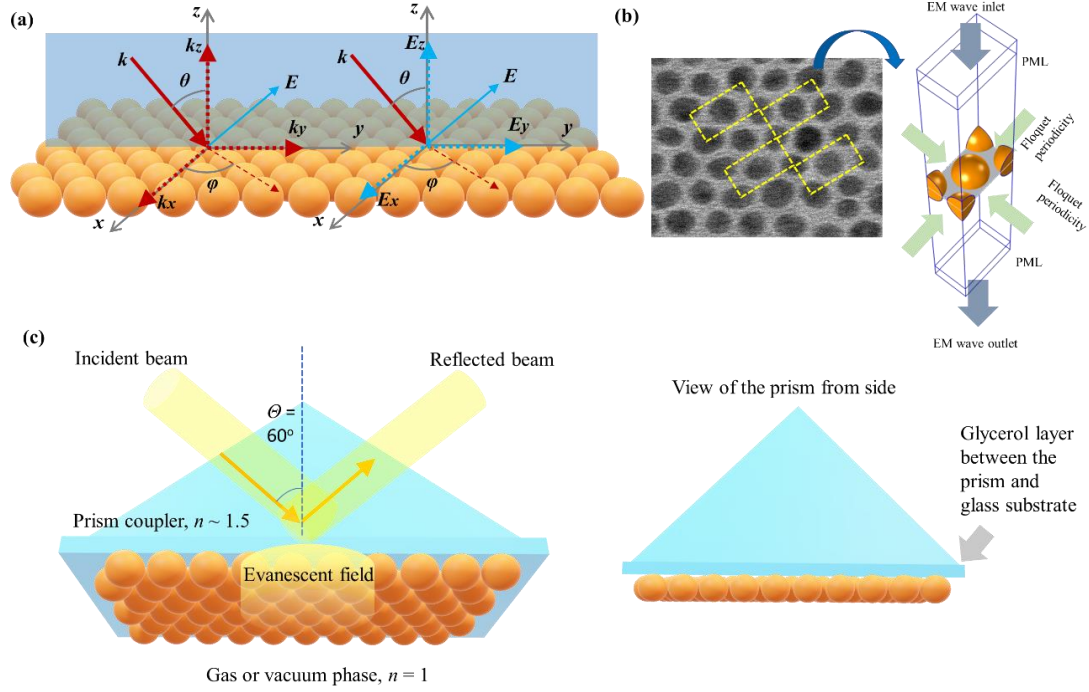

**Figure 2**. (a) Schematic representation of a Au nanoparticle self-assembled monolayer with incident and electric field wave vectors *w.r.t.* the 3-D cartesian axes and the plane of incidence (transparent blue plane). (b) The unit cell rectangular domain which is implemented in the electromagnetic modeling of the self-assembled films. (c) Schematic of the Kretschmann configuration: The Au nanoparticle film on glass substrate is attached to a prism coupler with a connecting glycerol layer in-between.

**Characterization of the films and computational validation**

Transmission Electron Microscopy (TEM) measurements were performed to obtain the spatial information required for electromagnetic modeling. As shown in **Figures 3 (a)** and **(b)**, a well-ordered, close-packed monolayer film is observed in TEM when nanoparticle concentrations were large enough to cover the complete air-liquid interface. When the concentration is too high, bilayer or multilayers form over significant areas of the film, **Figure 3 (d)** and **(e).** The nanoparticle size (~9.9 nm) and inter-particle gaps (~ 2 nm) are given in **Figure S5**. In colloids of such small nanoparticles, the Localized Surface Plasmon Resonance (LSPR) position and the bandwidth remain almost unaffected by the polydispersity of the particle diameter.[44] Schatz and co-workers have shown that irregularities in the 2D lattice, such as small variations in the particle diameter and the interparticle gap, can be circumvented in the electromagnetic models by taking average values, in our case 10 nm and 2 nm, respectively.[45] Also, minor deviations from the perfect hexagonal close packing can be ignored. Experimental and modeled transmittance spectra of the monolayer and bi-layer Au nanoparticle films (measurements done after vacuum drying at 40 °C for 72 hours) are compared in **Figures 3(c)** and **(f)**, respectively. The agreement between the experimental and theoretical resonance wavelength is clear from the position of the dip at 585 nm. Also in bilayer films, the relative position of the two nanoparticle layers has only a weak effect on the optical spectra, as shown in **Figure S2**. The film's collective plasmon resonance position is significantly red-shifted *w.r.t.* the resonance of isolated nanoparticles (521 nm, *cf.* **Figure 3 (i)**) due to the interparticle coupling. Importantly, these ~10 nm nanoparticles almost do not scatter light when isolated. However, as a close packed film, coherent scattering due to plasmonic coupling occurs, this phenomenon results in a significant reflectance (**Figure S7**). As small nanoparticles exhibit only non-radiative (or non-scattering) damping of the plasmons, the excited plasmons couple by near-field interactions among adjacent nanoparticles resulting in the "non-diffractive" lattice resonances. Due to their non-radiative property, the 10 nm Au nanoparticles cannot couple by far-field or radiative coupling. As shown in **Figures 3 (g)** and **(h),** for an interparticle gap of 2 nm, the near-field coupling is quite strong as indicated by the field enhancement maps. This strong near-field enhancement of up to 18 and 12 times the energy of the incident field for the monolayer and the bilayer, respectively, implies the applicability of these films in surface-enhanced processes such as SERS, SEIRA, Fluorescence, *etc*. Despite the same interparticle distance, the weaker near-field enhancement for the bi-layer film compared to the monolayer is due to the fact that the wave transmitted through the top layer is reflected back by the bottom layer to destructively interfere with the top layer field.[1]

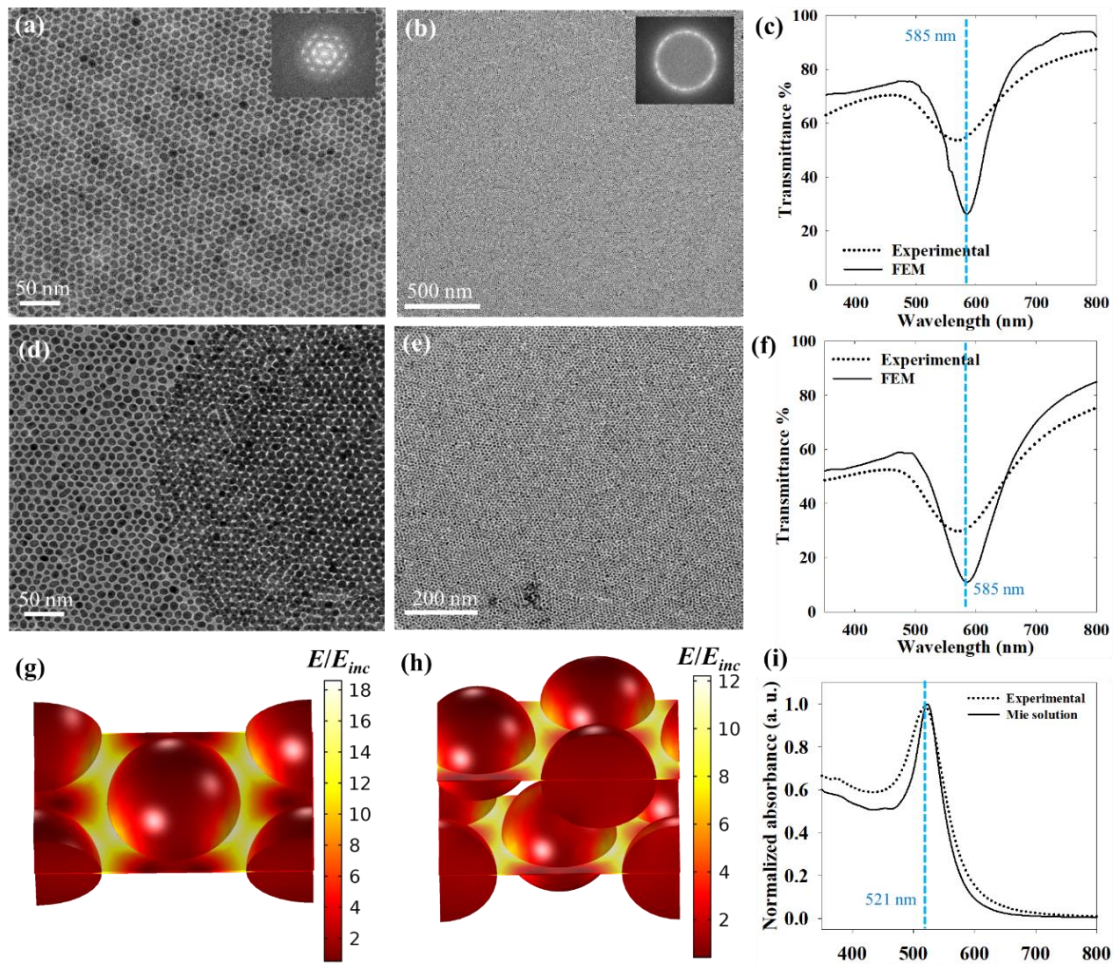

**Figure 3**. (a, b) TEM images of a Au nanoparticle monolayer with fast Fourier transformation (FFT) in the inset. (c) Comparison of experimental and computed transmission spectra of the Au nanoparticle monolayer. (d, e) TEM images of a respective Au nanoparticle bilayer. In (d), the contrast between bilayer and monolayer is visible. (f) Comparison of experimental and computed transmission spectra of the Au nanoparticle bilayer. (g, h) Near-field enhancement at the resonance of monolayer and bilayer films *w.r.t.* the incident field. (i) UV-Vis absorbance spectra of the Au nanoparticles as aqueous colloids compared with the Mie analytical spectrum for spherical Au nanoparticles (10 nm).

**Refractive index sensitivity of the optical response**

For the optical characterization and further sensing experiments, a nanoparticle film as shown in **Figure 1** was transferred to a 9 cm² glass slide. A large film area with uniform morphology is beneficial for the sensitivity of these measurements so that the entire incident beam during UV-Vis spectroscopy or ellipsometry can interact with the plasmonic nanoparticle film. Absorbance, reflectance, and transmittance spectra for normal incidence from the top were collected to show the agreement with the predicted plasmon resonance (**Figure 4 (a)** and **(b)**). Individually non-scattering ~10 nm Au nanoparticles packed as a film exhibit strong reflection by coherent scattering at the collective plasmon resonance of the lattice. The film's theoretical refractive index sensitivity of 250 nm/RIU in the normal incidence indicated by the red-shift of the collective resonance (*cfr.* optical intensity spectra) is expectedly similar to the sensitivity of a grating coupler, **Figure 4 (c)**.[46] Similar sensitivities have also been reported for the nanoparticle-on-a-mirror (NPOM) configuration[47] or nanoparticle films embedded in silica matrix.[48] It is important to note that the refractive index of the entire embedding medium outside of the prism is varied in the computations. In the Kretschmann configuration, the evanescent wave outside the prism coupler excites the plasmon polaritons in the nanoparticle film. The evanescent wave from *s*-polarized incidence attenuates faster with the distance from the interface than the evanescent field from *p*-polarized incidence.[49] In **Figure 4 (d, e),** the computed optical responses of the Au nanoparticle film to *p*- and *s*-polarized incident beams are significantly. The reflectance dip in both cases indicates the collective plasmon resonance of the film. However, the resonance band is weaker for the *s*-polarized beam as the penetration depth of the evanescent field for an *s*-polarized incident beam is smaller.[50] The polarization of the evanescent field due to *s*-polarized incident light is confined to the plane of the dielectric interface. In contrast, the evanescent field by *p*-polarized light has both in-plane and out-of-plane components. These results in the differences in the optical response of the two types of polarization. Regardless of the polarization, the refractive index sensitivity of the film in the Kretschmann configuration in **Figure 4 (d, e)** is similar to that of the normal incidence in **Figure 4 (c)**. Generally, in SPR- or LSPR-based sensing, the shift in the plasmon resonance wavelength or angle is correlated with the refractive index change induced by the target molecules in the surrounding environment. In the optical intensity mode as discussed so far, the wavelength interrogation in the Kretschmann configuration has been shown in literature to be more sensitive compared to the direct incidence configuration for nanostructured arrays such as diffraction gratings.[51] The differences in the polarization of the nanoparticles by *p*- and *s*-polarized incident beams are evident from the near-field maps in **Figure 4 (h)**. The ellipsometric responses originating from this polarization difference in optical response, is sensitive to the changes in the dielectric properties in the direct surroundings of the surface. The $\psi$ spectra, corresponding to the amplitude ratio of reflected *s*- and *p*-polarized components, are obtained from the complex reflectivity values for the two directions of polarization. They indicate the collective plasmon resonance of the film by a characteristic dip in the $\psi$ spectra, as seen in **Figure 4 (f)**, in line with the intensities in **Figure 4 (d)** and **4 (e)**. The Δ spectra, corresponding to the phase difference between the reflected *s*- and *p*-polarized components, in **Figure 4 (g)**, show a step close to the resonant wavelength. The theoretical refractive index sensitivity study of the amplitude ratio $\psi$ shows that similar to intensity spectra, increasing refractive index of the air (or vacuum) medium results in a red-shift of the plasmon band with an enhancement in the absolute value. The Δ spectra also show a similar response to refractive index changes, however, with >5 times higher sensitivity than $\psi$. Moirangthem *et al.* showed similar experimental trends in $\psi$ and Δ spectra in the liquid phase over a larger refractive index range in the sensing of Bovine serum albumin (BSA).[52] In the computational models for **Figure 4 (c)** to **4 (g)**, the refractive index change is considered homogeneous around the nanoparticles in the interparticle gaps as well as the rest of the medium. In reality, the probe molecules adsorb around the nanoparticles inhomogeneously, giving rise to further differences in the *p*- and *s*-polarized responses and hence, determining the true ellipsometric response. It has been shown that for a thin Au film, the $\psi$ and Δ responses from ellipsometry are 10 times more sensitive than in conventional intensity-based spectroscopy.[53] A similar trend is also observed in the wavelength interrogation with orders of magnitude higher shift in Δ spectra as compared to intensity spectra, shown in later sections when applied to methanol vapor sensing.

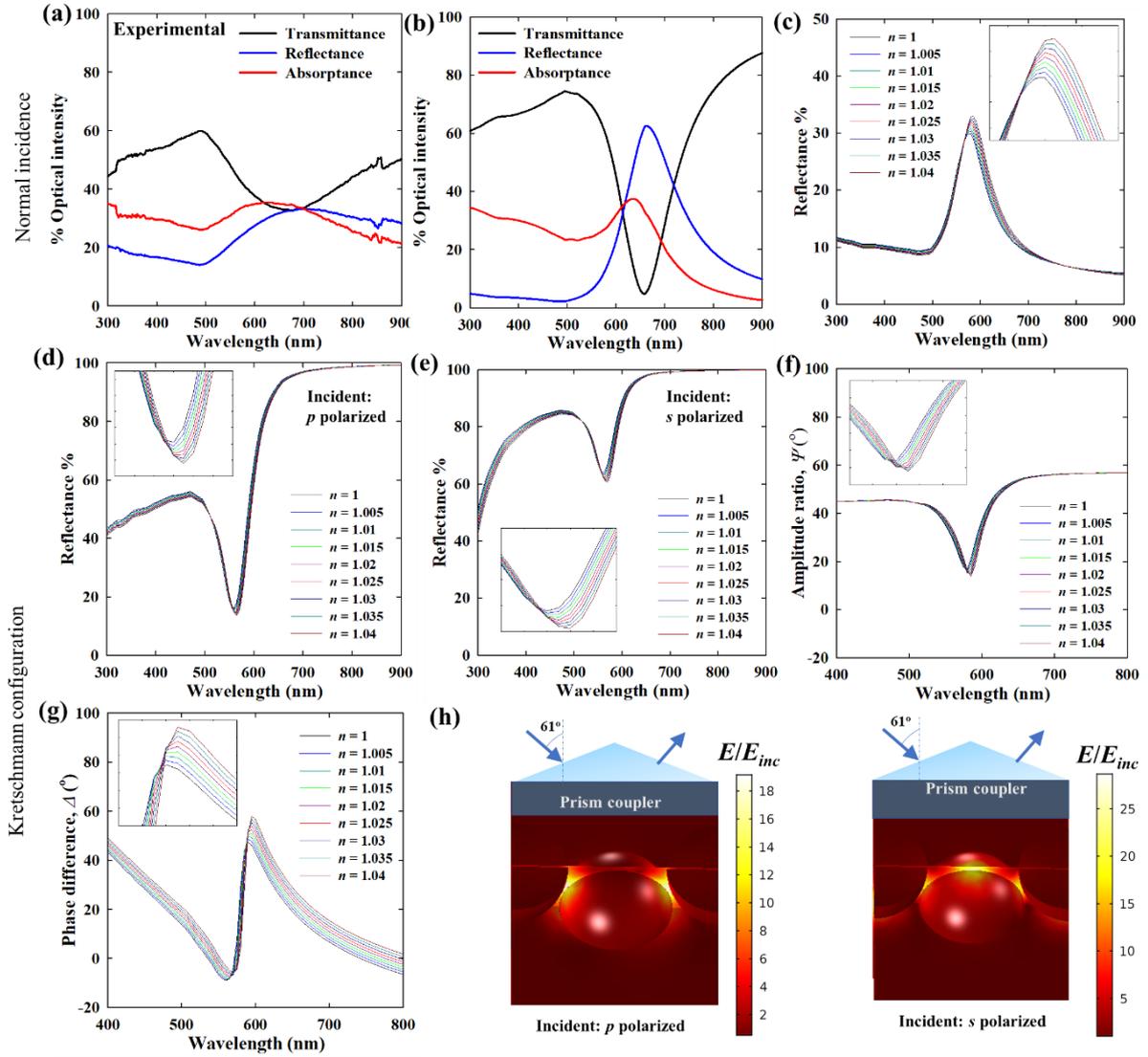

**Figure 4**. (a) Experimental transmittance, reflectance, and absorptance spectra of a self-assembled film of ~10 nm Au nanoparticles. (b) Computed optical spectra of the films for comparison. (c) Computed refractive index sensitivity of the Au nanoparticle film used in the sensing experiment under normal incidence. Computed reflectance spectra and refractive index sensitivity of *p*-polarized incident light (d) and *s*-polarized incident light (e) in the Kretschmann (*i.e.* ATR) configuration. Computed (f) $\psi$ and (g) $\Delta$ spectra from ellipsometry and their refractive index sensitivity for the Au nanoparticle film in the Kretschmann configuration. (h) Near-field enhancement around the self-assembled nanoparticles by *p*- and *s*-polarized incident beam in the Kretschmann configuration.

It is clear that 9 to 10 nm nanoparticles can couple strongly to exhibit a strong plasmon resonance of the lattice for sensing applications. With smaller sizes < 5 nm, the quantum size effects may become too strong weakening the individual plasmonic response as well as the coupling.[54] With larger nanoparticles, the coupling can be stronger and larger inter-particle voids can facilitate more adsorption/absorption of probe molecules. However, with larger nanoparticles *i.e.* larger surface area, larger number of probe molecules will also be required to have a measurable effect on the optical signal. Thus, investigations for an optimal size is interesting for future investigations.

*In-situ* **VOC (methanol) sensing**

A major advantage of periodic plasmonic nanostructure arrays (*e.g.* Au nanoparticle films) over traditional planar thin film SPR sensors for refractive index sensing is the highly localized near-field enhancement (**Figure 3(g), 4(h)**). The sensor response is only sensitive to the changes in the dielectric properties close or in to the Au nanoparticle film. A probe molecule can be selectively detected from a mixture if the molecules are selectively adsorbed on the nanoparticles' surface. Even though the aspect of selectivity is outside the scope of this work, the *in-situ* sensing experiments here demonstrate the refractive index-based sensing of vapor phase methanol molecules ($N_2$ gas and methanol mixture) by two optical configurations as discussed below.

***Methanol sensing experiments with in-situ spectroscopy***. Intensity-based spectroscopy at direct normal incidence is one of the simplest optical sensing set-ups. In this configuration, the choice of reflectance or transmittance is equivalent as the plasmon band in the spectra in both cases originates from the same collective plasmon resonance. It is also shown in Figure S8 that the ligand (*i.e.* dielectric shell around the nanoparticles) does not result in any significant changes in the sensitivity to the refractive index in the optical response. Next, the methanol dosing experiments were set up in which the refractive index in the bulk gas phase and in the vicinity of the nanoparticles was systematically varied while the film was continuously monitored at normal incidence in transmission mode, **Figure 5 (a)**. The refractive index change in the vicinity of the nanoparticles is induced by the adsorption of the methanol molecules on the ligand-nanoparticle network. As shown in **Figure 5 (b)**, the red-shift of the plasmon resonance is too small to be quantified accurately. However, the intensity changes can be exploited to track the refractive index changes over time as shown in **Figure 5 (c)** and (d). The sensitivity and reversibility of the signal are illustrated in the dosing experiment ($P/P_0$ from 0% to 65.58%) in **Figure 5 (c)** and **5 (d)**. The sensitivity (with reversibility) up to 4.63% methanol vapor indicates that methanol adsorption in the organic ligand network contributes strongly to the refractive index increase in the vicinity of the Au nanoparticles. Also, **Figure 3** shows that the evanescent near-field is confined close to the nanoparticles and in the interparticle gaps. Thus, the near-field can only interact with the local dielectric environment over a short-range, therefore the far-field response also mainly represents sensitivity in these regions.

to the total internal reflection. At 60° incidence angle, the reflectance for *p*- and *s*-polarized incidence show a clear plasmon band. The

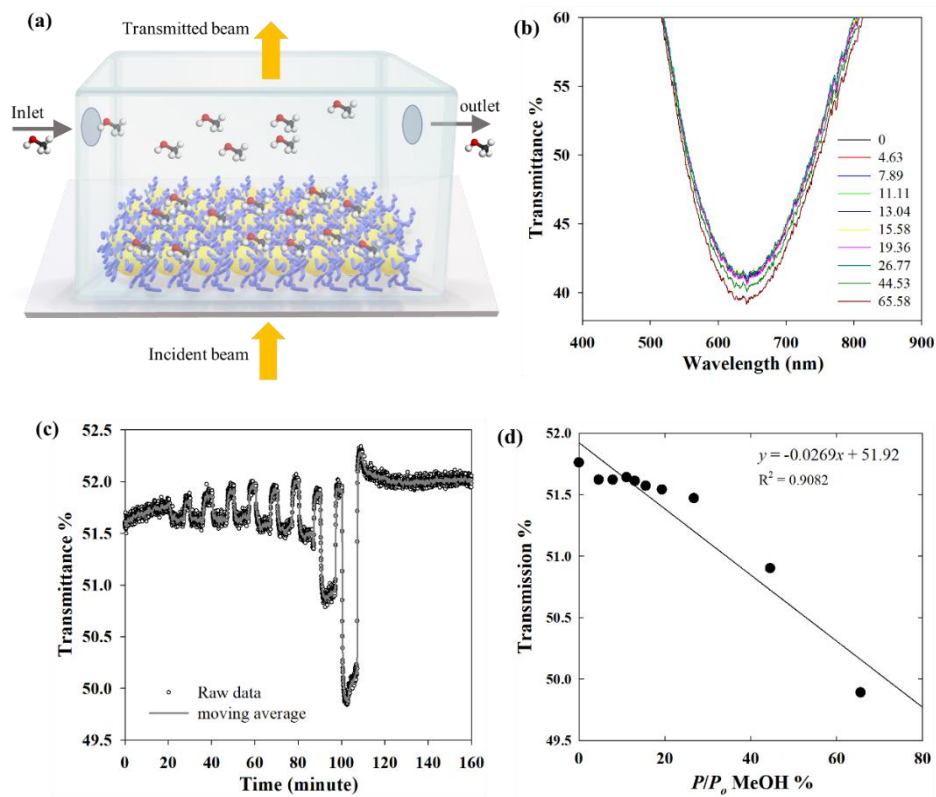

**Figure 5**. (a) Methanol vapor dosing set-up with in-situ transmission spectroscopy (yellow spheres: Au nanoparticles, blue wires: oleylamine ligand). (b) Transmittance at the resonance wavelength for increasing relative methanol saturation (% $P/P_o$). (c) Time evolution of the transmittance at wavelength: 650 nm during the methanol dosing experiment with increasing concentration. (d) Isotherm showing the dependence of the % transmittance *vs*. methanol %. Nanopartices are shown not to scale. (incidence angle for the Kretschmann configuration: 61°).

*Methanol sensing in the Kretschmann (ATR) configuration with in-situ ellipsometry*. The Kretschmann configuration as shown in **Figure 2** is exploited in sensing applications due to several advantages. Firstly, in contrast to direct incidence, the attenuated total internal reflection (ATR) mode preserves all the light that is not absorbed by the sample itself, *i.e.* no transmittance loss in reflectance. Secondly, as the probe beam remains outside the setup, only the evanescent beam interacts with the sample. Thus, the presence of contaminants or device components in the bulk does not interfere in the optical path. The *in-situ* methanol sensing experiments in the Kretschmann configuration were set-up with an angle of incidence of 61°. **Figure S10** shows the angular dependence of optical spectra in the Kretschmann configuration. For incidence angles above the critical angle (~41°), the transmittance is 0 due difference in the reflectance by the two polarization gives rise to the ellipsometric response in $\psi$.

The use of ellipsometry instead of intensity-based probing enables exploitation of polarization-specific amplitude (as the amplitude ratio, $\psi$) and phase information (as the phase difference, $\Delta$) of the film. By exploiting the polarization dependence of the plasmonic response to refractive index changes in the nanostructure surrounding, higher sensitivity can be obtained when using $\psi$ and $\Delta$.[55] In the methanol vapor dosing experiments, in contrast to the computational models where the refractive index is varied uniformly in the entire medium, the local refractive index changes due to molecular adsorption in the ligand-nanoparticle network is not necessarily uniform. **Figure 6 (a)** shows the open side of the film facing the gas phase, which is expected to naturally contain a higher concentration of the methanol molecules, compared to the interstitial gaps that pose an additional diffusion resistance. The non-uniform distribution of methanol adsorption in the vicinity of the nanoparticles is also expected to contribute to the difference in the optical response of the two orthogonal modes of incident polarization with differently directed evanescent fields. In the Kretschmann configuration, it has been shown that while the *p*-polarized incident beam can probe the vibration modes parallel and perpendicular to the sample film, the *s*-polarized beam probes solely the components of the vibrational modes parallel to the plane of the sample film.[56] The *in-situ* ellipsometry set-up in this work enables to elucidate the effect of the adsorption of methanol molecules on the plasmonic nanoparticle-organic ligand network by means of the relative optical response to the in-plane and out-of-plane polarized evanescent fields. In the set-up, the self-assembled Au nanoparticle film is combined with a (bottom face area: 4 cm²) prism coupler to construct the Kretschmann configuration inside an environmental chamber into which the methanol vapor was dosed systematically, **Figure 6 (a)**.

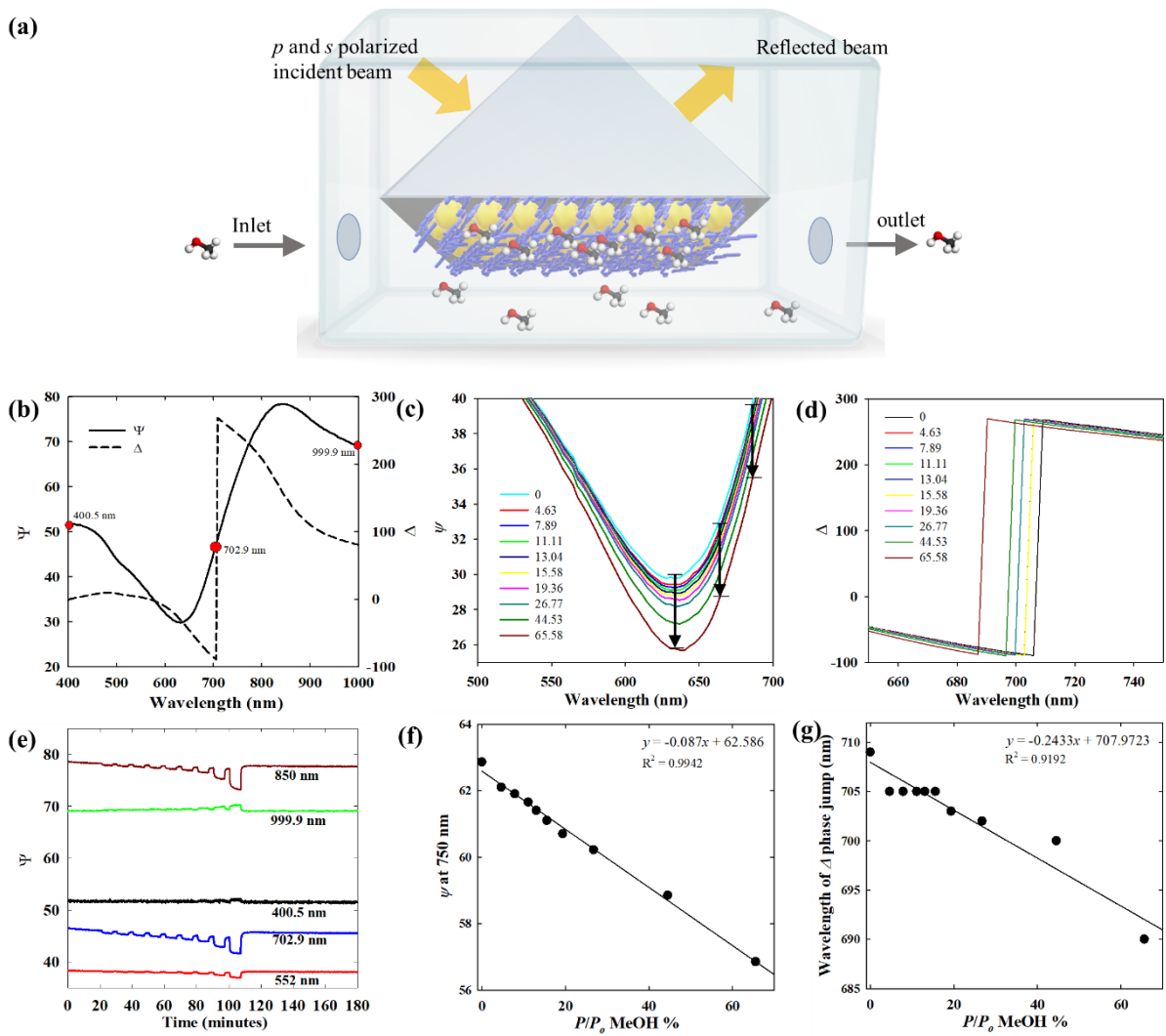

**Figure 6**. (a) The in-situ ellipsometry set-up used for the methanol dosing experiment. (b) Experimental ellipsometric $\psi$ and $\Delta$ spectra at 61° incidence angle. (c) $\psi$ at the resonance wavelength for increasing relative methanol saturation (% $P/P_o$). (d) Time evolution of $\psi$ during the methanol dosing experiment at different wavelengths. (e) Isotherm showing the dependence of the $\psi$ vs. methanol %. (f) Isotherm exported from $\psi$ intensity. (g) Isotherm obtained from wavelength position of phase $\Delta$ jump.

**Figure 6 (b)** shows the $\psi$ and $\Delta$ spectra of the nanoparticle film in the Kretschmann geometry.

The dip in $\psi$ around ~640 nm again represents the plasmon resonance. In the $\Delta$, a corresponding jump at ~700 nm is observed. The characteristic $\psi$ and $\Delta$ spectra of the film as shown in **Figure 6 (b)** agree well with the computed spectra in **Figure 4**. The variation of $\psi$ and $\Delta$ with increasing concentration of methanol vapor indicates the changes in the refractive index close to the nanoparticle film, **Figure 6 (c, d)**. Although the red-shift of the plasmon band in the $\psi$ spectra is rather weak, the intensity experiences a significant variation with methanol vapor concentration similar to the computed trends in **Figure 4 (f).** In contrast, $\Delta$ shows a strong red-shift of ~20 nm for the methanol vapor concentration variation from 0 to 65.58 % ($P/P_o$). Due to the non-uniform refractive index change throughout the ligand-nanoparticle matrix, the experimental $\psi$ and $\Delta$ spectra variation with different methanol concentrations is expectedly different from the computed results shown for the model nanoparticle film. Still, the general agreement between the experiments in **Figure 6** and the models in **Figure 4** is clear with respect to the characteristic plasmon band in $\psi$ and the jump in $\Delta$. **Figure 4** also shows how the near-field enhancement of the film is distributed and the sensitivity to refractive index changes is strongly dependent on the direction of polarization.[57] **Figure 6 (e)** shows the intensity variation of the dip in the $\psi$ spectra around 640 nm with increasing methanol vapor concentration. Probing the intensity at wavelengths longer than the resonance wavelength is also useful as the change in $\psi$ with concentration is of a similar order of magnitude. The time evolution of the $\psi$ intensity in **Figure 6 (e)** shows the reversibility of the equilibrium adsorption/desorption process. Compared to the transmission mode experiments, an improvement in the noise can be observed, especially at wavelengths close to the resonance. The sensitivity is the highest close to the resonance at ~ 640 nm which gradually decreases as one moves away from the resonance wavelength. The concentration vs. $\psi$ data in **Figure 6 (f)** indicates a linear relationship. A similar linear trend is observed in the concentration vs. $\Delta$ position data in **Figure 6 (g),** the clarity of which however suffers from the wavelength resolution of the ellipsometer. The concentration vs. $\Delta$ position data in **Figure 6 (g)** show that the high sensitivity of the phase difference $\Delta$ can be exploited to obtain at least an order of magnitude higher sensitivity than $\psi$-based probing. While the collective resonances in 2D Au nanostructure arrays can also be excited by normal incidence, the Kretschmann configuration coupled to ellipsometric measurements is particularly advantageous. We hypothesize that the improved noise level can be attributed to the fact that the plasmonic excitation in the Kretschmann geometry is driven by the evanescent wave, while the incident beam is reflected by total internal reflection with no transmission. Thus, unlike the transmission mode where reflection losses are significant, the reflectance in the Kretschmann configuration facilitates the collection of the entire light signal that is not absorbed by the nanoparticle film upon plasmonic excitation. Also, $\psi$ and $\Delta$ being relative parameters are free from the potential noise arising from the fluctuations in the light source.

It is important to note that the ligand-nanoparticle network does not have exceptionally high surface area to accommodate a large number of probe molecules. Furthermore, no specific surface chemistry is incorporated for the adsorption selectivity. However, the selectivity in certain cases can be determined by the hydrophobicity (or polarity) of the film. In this regard, combining a metal nanoparticles with a porous material, such as metal-organic frameworks (MOFs), can be a promising approach to increase both the adsorption capacity as the selectivity.[9,58,59,60,61,62]

**Conclusion**

Self-assembly of nanoparticles is a simple method to obtain functional thin films. At the ethylene glycol-air interface, colloidal nanoparticles can be self-assembled over large areas up to several cm$^2$ by controlled evaporation of the solvent phase. Once self-assembled into close-packed films, oleylamine capped plasmonic Au nanoparticles exhibit collective optical properties through plasmonic coupling. Such nanoparticle films exhibit a strong "non-diffractive" lattice resonance by the near-field coupling, as shown by electromagnetic modeling. Due to the periodicity in the nanoparticle arrangement, the collective plasmon modes of self-assembled Au nanoparticle films can be excited by normally incident electromagnetic waves. However, the collective coupled plasmon resonance can also be excited in the Kretschmann configuration by the interaction of the evanescent field with the nanoparticles at the low refractive index side. The refractive index sensitivity of the intensity spectra (*i.e.*, transmittance, absorptance, or reflectance) of the self-assembled films in both normal incidence and Kretschmann configuration is similar to grating-based plasmonic sensing platforms reported in literature. When *in-situ* ellipsometry and *in-situ* UV-Vis spectroscopy are compared, amplitude ratio ($\psi$) and phase difference ($\Delta$) measurements from ellipsometry provide significantly improved sensitivity, especially when the phase difference ($\Delta$) is probed.

The major advantage of using a periodically structured nanoparticle film as a sensing platform, is that the optical response can be highly selective only to the molecules that are adsorbed on nanoparticles. This, combined with the advantages of the Kretschmann configuration and ellipsometry, can be promising for highly sensitive sensors. As a next step, further functionalization of these nanoparticle films is encouraged to enable selective and highly sensitive sensing of volatile organic compounds (VOCs).

**Methods**

*Synthesis of Au nanoparticles and phase transfer.* Au nanoparticles were synthesized by the citrate reduction method. A 100 mL 5 mM tri-sodium citrate solution in water was brought to a boil before the addition of 1 mL of 25 mM $HAuCl_4$ solution. Following the precursor addition, the solution was kept boiling for 20 more minutes for complete reduction. The as-synthesized nanoparticles were centrifuged at 11529g for 30 minutes and redispersed in 10 mL Milli-Q water to obtain a ten times more concentrated colloid.

The concentrated Au nanoparticles in the aqueous phase were transferred to the toluene phase by using oleylamine as a hydrophobizing ligand. 3 mL of toluene were added onto the surface of 1 mL of aqueous Au nanoparticle colloid suspension, so that the toluene floats over the water as a separate phase. To this bi-phasic liquid, 2 mL of oleylamine in ethanol solution (200 mg in 20 mL ethanol) were added followed by gentle shaking for mixing and then resting for separation of the toluene and ethanol-water phases. After resting the mixture for 24 hours, the nanoparticles appear in the toluene phase above (indicated by a distict red color) and the ethanol-water phase below becomes transparent. The nanoparticles in the toluene phase were then carefully pipetted out for further processing.

*Self-assembly.* The nanoparticles in toluene (4 mL) after phase transfer were concentrated 10 times to 400 µl by centrifugation at 10625g for 30 minutes and then introduced onto the surface of ethylene glycol contained in a 40 mL glass beaker of 4 cm in diameter. The beaker was then immediately covered with a glass plate and parafilm, leaving only a tiny hole at the mouth of the beaker for the toluene to evaporate slowly. After the complete evaporation of toluene (*ca*. 8 hours), the nanoparticle films were transferred to a glass substrate by careful vertical dip coating similar to the Langmuir-Blodgett technique. The films were vacuum dried at 100°C for 24 hours to remove the residual ethylene glycol. For bi-layer films, two films were deposited sequentially one on the other.

*Characterization.* The UV-Vis transmittance spectra for the model validation of the initial films were taken by a Shimadzu UV 2600 spectrophotometer. The reflectance and transmittance of the film used in the *in-situ* measurements were taken by a Lambda UV/VIS 950 Perkin Elmer Spectrophotometer. The absorbance was calculated from the measured reflectance and transmittance (100% - R% - T% = A%). Bright-field transmission electron microscopy (TEM) images of self-assembled nanoparticles were acquired using a Thermo Fisher Scientific Tecnai G2 operated at an accelerating voltage of 200 kV. The self-assembled film was immobilized on a hydrophobic copper grid using the same procedure to immobilize self-assembly on glass substrates. The grid was dried at 100°C for 12 h in a vacuum oven. Scanning electron microscopy (SEM) images of the films were acquired using an FEG-ESEM-EDX, Thermo Fisher Scientific Quanta 250 at an accelerating voltage of 20 kV.

*Dosing experiments.* Vapor concentrations of methanol in $N_2$ flow were regulated in a custom-built dosing setup. Concentrations were set by controlling the liquid methanol and $N_2$ gas flow with mass flow controllers. The methanol was evaporated and mixed with the $N_2$ in a controlled evaporation mixer. The methanol and $N_2$ gas mixture was then passed through an environmental chamber where the Au nanoparticle film was set-up for *in-situ* spectroscopic and ellipsometric measurements. During each run, different concentrations were swept, alternated by a purge step to assess the reversibility of the response.

*Conventional normal incidence configuration with in-situ spectroscopy.* The nanoparticle film on the glass substrate was placed on the sample holder, resulting in an upwards-facing gold nanoparticle film on which the incident beam was directed from the bottom (simplified schematic in **Figure 5**). An Avantes fiber optic UV-VIS-NIR spectrometer and an Avalight balanced deuterium light source were used in the *in-situ* set-up. Measurements were performed in the transmission mode using Thorlabs optical and mechanical components. A linkam THMS600 was used as an environmental cell.

*Kretschmann configuration with in-situ ellipsometry.* The Au nanoparticle film on the glass substrate was placed facing the sample holder, using thin spacers at the edges to avoid contact and allow gas flow to pass the particles. A right-angle prism is placed on the top of the glass slide using glycerol as a contact fluid. The beam of the ellipsometer is let in/out through fused silica glass windows of the stainless-steel environmental chamber that contains the set-up. All ellipsometry measurements were performed with an iSE Woollam Ellipsometer at an angle of 61° from the normal.

*Electromagnetic modeling.* For the electromagnetic modeling of the nanoparticle films, a Finite Element Method (FEM) solver in COMSOL Multiphysics was used to solve the frequency domain form of the Maxwell's equations numerically. The infinite 2D films were approximated with rectangular unit cells with periodic boundary conditions on the side walls. The incident electromagnetic field was excited at the top as shown in **Figure 2 (b)** (and the wave outlet at the bottom) and the discretized equations were solved for the full field solution. A perfectly matched layer (PML) was implemented on the top and the bottom boundaries for the complete absorption of the propagating wave. The mesh and the domain independence of the numerical results were ascertained by performing mesh and domain independence tests, and comparison with existing numerical/experimental results. More details/schematics about the numerical results are provided in the result and discussion section, and **Figure S1**. (further information in the supporting document)

## Author Contributions

R.B. and J.S. conceived the study, performed the experiments, modeling and wrote the manuscript. R.N. performed the TEM analysis. M.L.T. contributed to the ellipsometry measurements. D.N.C. supervised the electromagnetic modeling. S.B. supervised the TEM analysis. S.L., R.A. and S.W.V. supervised the entire study, procured the funding and edited the final version of the manuscript.

## Supporting Information

Details of computational modeling, computational validation, particle size characterization, additional SEM and TEM characterization, refractive index/incidence angle dependence of optical spectra, MATLAB code for color simulation from optical spectra.

## Acknowledgment

R.B. acknowledges financial support from the University of Antwerp Special Research Fund (BOF) for a DOCPRO4 doctoral scholarship. J.S. acknowledges financial support from the Research Foundation Flanders (FWO) by a PhD fellowship (11H8121N). M.L.T. acknowledges financial support from the Research Foundation Flanders (FWO) by a senior postdoctoral fellowship (12ZK720N).

## References

(1) Kravets, V. G.; Kabashin, A. V.; Barnes, W. L.; Grigorenko, A. N. Plasmonic Surface Lattice Resonances: A Review of Properties and Applications. *Chem. Rev.* **2018**, *118* (12), 5912–5951. https://doi.org/10.1021/acs.chemrev.8b00243.

(2) Deng, T.-S.; Parker, J.; Hirai, Y.; Shepherd, N.; Yabu, H.; Scherer, N. F. Designing "Metamolecules" for Photonic Function: Reduced Backscattering. *physica status solidi (b)*

**2020**, *257* (12), 2000169. https://doi.org/10.1002/pssb.202000169.

(3) Yang, K.; Yao, X.; Liu, B.; Ren, B. Metallic Plasmonic Array Structures: Principles, Fabrications, Properties, and Applications. *Advanced Materials* **2021**, *33* (50), 2007988. https://doi.org/10.1002/adma.202007988.

(4) Stratakis, E.; Kymakis, E. Nanoparticle-Based Plasmonic Organic Photovoltaic Devices. *Materials Today* **2013**, *16* (4), 133–146. https://doi.org/10.1016/j.mattod.2013.04.006.

(5) Lim, S. Y.; Law, C. S.; Bertó-Roselló, F.; Liu, L.; Markovic, M.; Ferré-Borrull, J.; Abell, A. D.; Voelcker, N. H.; Marsal, L. F.; Santos, A. Tailor-Engineered Plasmonic Single-Lattices: Harnessing Localized Surface Plasmon Resonances for Visible-NIR Light-Enhanced Photocatalysis. *Catal. Sci. Technol.* **2020**, *10* (10), 3195–3211. https://doi.org/10.1039/C9CY02561H.

(6) Song, L.; Qiu, N.; Huang, Y.; Cheng, Q.; Yang, Y.; Lin, H.; Su, F.; Chen, T. Macroscopic Orientational Gold Nanorods Monolayer Film with Excellent Photothermal Anticounterfeiting Performance. *Advanced Optical Materials* **2020**, *8* (18), 1902082. https://doi.org/10.1002/adom.201902082.

(7) Wang, H.; Yao, L.; Mao, X.; Wang, K.; Zhu, L.; Zhu, J. Gold Nanoparticle Superlattice Monolayer with Tunable Interparticle Gap for Surface-Enhanced Raman Spectroscopy. *Nanoscale* **2019**, *11* (29), 13917–13923. https://doi.org/10.1039/C9NR03590G.

(8) Hirai, Y.; Matsuo, Y.; Yabu, H. Near-Infrared-Excitable SERS Measurement Using Magneto-Responsive Metafluids for in Situ Molecular Analysis. *ACS Appl. Nano Mater.* **2018**, *1* (9), 4980–4987. https://doi.org/10.1021/acsanm.8b01093.

(9) Qiao, X.; Su, B.; Liu, C.; Song, Q.; Luo, D.; Mo, G.; Wang, T. Selective Surface Enhanced Raman Scattering for Quantitative Detection of Lung Cancer Biomarkers in Superparticle@MOF Structure. *Advanced Materials* **2018**, *30* (5), 1702275. https://doi.org/10.1002/adma.201702275.

(10) Pang, J. S.; Theodorou, I. G.; Centeno, A.; Petrov, P. K.; Alford, N. M.; Ryan, M. P.; Xie, F. Tunable Three-Dimensional Plasmonic Arrays for Large Near-Infrared Fluorescence Enhancement. *ACS Appl. Mater. Interfaces* **2019**, *11* (26), 23083–23092. https://doi.org/10.1021/acsami.9b08802.

(11) Mayerhöfer, T. G.; Popp, J. Periodic Array-Based Substrates for Surface-Enhanced Infrared Spectroscopy. *Nanophotonics* **2018**, *7* (1), 39–79. https://doi.org/10.1515/nanoph-2017-0005.

(12) Piltan, S.; Sievenpiper, D. Plasmonic Nano-Arrays for Enhanced Photoemission and Photodetection. *J. Opt. Soc. Am. B, JOSAB* **2018**, *35* (2), 208–213. https://doi.org/10.1364/JOSAB.35.000208.

(13) Grzelczak, M.; Vermant, J.; Furst, E. M.; Liz-Marzán, L. M. Directed Self-Assembly of Nanoparticles. *ACS Nano* **2010**, *4* (7), 3591–3605. https://doi.org/10.1021/nn100869j.

(14) Borah, R.; Ninakanti, R.; Nuyts, G.; Peeters, H.; Pedrazo-Tardajos, A.; Nuti, S.; Velde, C. V.; Wael, K. D.; Lenaerts, S.; Bals, S.; Verbruggen, S. W. Selectivity in the Ligand Functionalization of Photocatalytic Metal Oxide Nanoparticles for Phase Transfer and Self-Assembly Applications. *Chemistry – A European Journal* **2021**, *27* (35), 9011–9021. https://doi.org/10.1002/chem.202100029.

(15) Thorkelsson, K.; Bai, P.; Xu, T. Self-Assembly and Applications of Anisotropic Nanomaterials: A Review. *Nano Today* **2015**, *10* (1), 48–66. https://doi.org/10.1016/j.nantod.2014.12.005.

(16) Lei, Y.; Chim, W.-K. Highly Ordered Arrays of Metal/Semiconductor Core−Shell Nanoparticles with Tunable Nanostructures and Photoluminescence. *J. Am. Chem. Soc.* **2005**, *127* (5), 1487–

1492. https://doi.org/10.1021/ja043969m.
(17) Zhu, H.; Fan, Z.; Yu, L.; Wilson, M. A.; Nagaoka, Y.; Eggert, D.; Cao, C.; Liu, Y.; Wei, Z.; Wang, X.; He, J.; Zhao, J.; Li, R.; Wang, Z.; Grünwald, M.; Chen, O. Controlling Nanoparticle Orientations in the Self-Assembly of Patchy Quantum Dot-Gold Heterostructural Nanocrystals. *J. Am. Chem. Soc.* **2019**, *141* (14), 6013–6021. https://doi.org/10.1021/jacs.9b01033.
(18) Deng, K.; Luo, Z.; Tan, L.; Quan, Z. Self-Assembly of Anisotropic Nanoparticles into Functional Superstructures. *Chem. Soc. Rev.* **2020**, *49* (16), 6002–6038. https://doi.org/10.1039/D0CS00541J.
(19) Song, L.; Xu, B. B.; Cheng, Q.; Wang, X.; Luo, X.; Chen, X.; Chen, T.; Huang, Y. Instant Interfacial Self-Assembly for Homogeneous Nanoparticle Monolayer Enabled Conformal "Lift-on" Thin Film Technology. *Science Advances* **2021**, *7* (52), eabk2852. https://doi.org/10.1126/sciadv.abk2852.
(20) Dong, A.; Chen, J.; Vora, P. M.; Kikkawa, J. M.; Murray, C. B. Binary Nanocrystal Superlattice Membranes Self-Assembled at the Liquid–Air Interface. *Nature* **2010**, *466* (7305), 474–477. https://doi.org/10.1038/nature09188.
(21) Elbert, K. C.; Zygmunt, W.; Vo, T.; Vara, C. M.; Rosen, D. J.; Krook, N. M.; Glotzer, S. C.; Murray, C. B. Anisotropic Nanocrystal Shape and Ligand Design for Co-Assembly. *Science Advances* **2021**. https://doi.org/10.1126/sciadv.abf9402.
(22) Kanie, K.; Matsubara, M.; Zeng, X.; Liu, F.; Ungar, G.; Nakamura, H.; Muramatsu, A. Simple Cubic Packing of Gold Nanoparticles through Rational Design of Their Dendrimeric Corona. *J. Am. Chem. Soc.* **2012**, *134* (2), 808–811. https://doi.org/10.1021/ja2095816.
(23) Lee, S.; Sim, K.; Moon, S. Y.; Choi, J.; Jeon, Y.; Nam, J.-M.; Park, S.-J. Controlled Assembly of Plasmonic Nanoparticles: From Static to Dynamic Nanostructures. *Advanced Materials* **2021**, *33* (46), 2007668. https://doi.org/10.1002/adma.202007668.
(24) Cersonsky, R. K.; Anders, G. van; Dodd, P. M.; Glotzer, S. C. Relevance of Packing to Colloidal Self-Assembly. *PNAS* **2018**, *115* (7), 1439–1444. https://doi.org/10.1073/pnas.1720139115.
(25) Lu, X.; Huang, Y.; Liu, B.; Zhang, L.; Song, L.; Zhang, J.; Zhang, A.; Chen, T. Light-Controlled Shrinkage of Large-Area Gold Nanoparticle Monolayer Film for Tunable SERS Activity. *Chem. Mater.* **2018**, *30* (6), 1989–1997. https://doi.org/10.1021/acs.chemmater.7b05176.
(26) Willets, K. A.; Van Duyne, R. P. Localized Surface Plasmon Resonance Spectroscopy and Sensing. *Annual Review of Physical Chemistry* **2007**, *58* (1), 267–297. https://doi.org/10.1146/annurev.physchem.58.032806.104607.
(27) Li, M.; Cushing, S. K.; Wu, N. Plasmon-Enhanced Optical Sensors: A Review. *Analyst* **2014**, *140* (2), 386–406. https://doi.org/10.1039/C4AN01079E.
(28) Nguyen, H. H.; Park, J.; Kang, S.; Kim, M. Surface Plasmon Resonance: A Versatile Technique for Biosensor Applications. *Sensors* **2015**, *15* (5), 10481–10510. https://doi.org/10.3390/s150510481.
(29) Kretschmann, E.; Raether, H. Notizen: Radiative Decay of Non Radiative Surface Plasmons Excited by Light. *Zeitschrift für Naturforschung A* **1968**, *23* (12), 2135–2136. https://doi.org/10.1515/zna-1968-1247.
(30) Schulz, F.; Pavelka, O.; Lehmkühler, F.; Westermeier, F.; Okamura, Y.; Mueller, N. S.; Reich, S.; Lange, H. Structural Order in Plasmonic Superlattices. *Nat Commun* **2020**, *11* (1), 3821. https://doi.org/10.1038/s41467-020-17632-4.
(31) Pohjalainen, E.; Pohjakallio, M.; Johans, C.; Kontturi, K.; Timonen, J. V. I.; Ikkala, O.; Ras, R. H. A.; Viitala, T.; Heino, M. T.; Seppälä, E. T. Cobalt Nanoparticle Langmuir−Schaefer Films on Ethylene Glycol Subphase.

*Langmuir* **2010**, *26* (17), 13937–13943. https://doi.org/10.1021/la101630q.

(32) Yu, J.; Zheng, L.; Geng, C.; Wang, X.; Yan, Q.; Wang, X.; Shen, G.; Shen, D. Colloidal Monolayer at the Air/Water Interface: Large-Area Self-Assembly and in-Situ Annealing. *Thin Solid Films* **2013**, *544*, 557–561. https://doi.org/10.1016/j.tsf.2012.12.069.

(33) Dai, Z.; Dai, H.; Zhou, Y.; Liu, D.; Duan, G.; Cai, W.; Li, Y. Monodispersed Nb2O5 Microspheres: Facile Synthesis, Air/Water Interfacial Self-Assembly, Nb2O5-Based Composite Films, and Their Selective NO2 Sensing. *Advanced Materials Interfaces* **2015**, *2* (11), 1500167. https://doi.org/10.1002/admi.201500167.

(34) Lau, C. Y.; Duan, H.; Wang, F.; He, C. B.; Low, H. Y.; Yang, J. K. W. Enhanced Ordering in Gold Nanoparticles Self-Assembly through Excess Free Ligands. *Langmuir* **2011**, *27* (7), 3355–3360. https://doi.org/10.1021/la104786z.

(35) Oliveira, O. N.; Caseli, L.; Ariga, K. The Past and the Future of Langmuir and Langmuir–Blodgett Films. *Chem. Rev.* **2022**, *122* (6), 6459–6513. https://doi.org/10.1021/acs.chemrev.1c00754.

(36) Wang, B.; Yu, P.; Wang, W.; Zhang, X.; Kuo, H.-C.; Xu, H.; Wang, Z. M. High-Q Plasmonic Resonances: Fundamentals and Applications. *Advanced Optical Materials* **2021**, *9* (7), 2001520. https://doi.org/10.1002/adom.202001520.

(37) Zhang, J.; Zhang, L.; Xu, W. Surface Plasmon Polaritons: Physics and Applications. *J. Phys. D: Appl. Phys.* **2012**, *45* (11), 113001. https://doi.org/10.1088/0022-3727/45/11/113001.

(38) Borah, R.; Verbruggen, S. W. Coupled Plasmon Modes in 2D Gold Nanoparticle Clusters and Their Effect on Local Temperature Control. *J. Phys. Chem. C* **2019**, *123* (50), 30594–30603. https://doi.org/10.1021/acs.jpcc.9b09048.

(39) Molet, P.; Passarelli, N.; Pérez, L. A.; Scarabelli, L.; Mihi, A. Engineering Plasmonic Colloidal Meta-Molecules for Tunable Photonic Supercrystals. *Advanced Optical Materials* **2021**, *9* (20), 2100761. https://doi.org/10.1002/adom.202100761.

(40) Sikdar, D.; Kornyshev, A. A. Theory of Tailorable Optical Response of Two-Dimensional Arrays of Plasmonic Nanoparticles at Dielectric Interfaces. *Sci Rep* **2016**, *6* (1), 33712. https://doi.org/10.1038/srep33712.

(41) Milosevic, M. On the Nature of the Evanescent Wave. *Appl Spectrosc* **2013**, *67* (2), 126–131. https://doi.org/10.1366/12-06707.

(42) Mueller, N. S.; Vieira, B. G. M.; Schulz, F.; Kusch, P.; Oddone, V.; Barros, E. B.; Lange, H.; Reich, S. Dark Interlayer Plasmons in Colloidal Gold Nanoparticle Bi- and Few-Layers. *ACS Photonics* **2018**, *5* (10), 3962–3969. https://doi.org/10.1021/acsphotonics.8b00898.

(43) Vohnsen, B.; Valente, D. Surface-Plasmon-Based Wavefront Sensing. *Optica, OPTICA* **2015**, *2* (12), 1024–1027. https://doi.org/10.1364/OPTICA.2.001024.

(44) Borah, R.; Verbruggen, S. W. Effect of Size Distribution, Skewness and Roughness on the Optical Properties of Colloidal Plasmonic Nanoparticles. *Colloids and Surfaces A: Physicochemical and Engineering Aspects* **2022**, *640*, 128521. https://doi.org/10.1016/j.colsurfa.2022.128521.

(45) Ross, M. B.; Ku, J. C.; Blaber, M. G.; Mirkin, C. A.; Schatz, G. C. Defect Tolerance and the Effect of Structural Inhomogeneity in Plasmonic DNA-Nanoparticle Superlattices. *PNAS* **2015**, *112* (33), 10292–10297. https://doi.org/10.1073/pnas.1513058112.

(46) Homola, J.; Koudela, I.; Yee, S. S. Surface Plasmon Resonance Sensors Based on Diffraction Gratings and Prism Couplers: Sensitivity Comparison. *Sensors and Actuators B:*

*Chemical* **1999**, *54* (1), 16–24. https://doi.org/10.1016/S0925-4005(98)00322-0.

(47) Zhu, Z.; Ding, Y.; Wang, Z.; Cheng, C.; Li, D.; Chen, H. High-Performance Plasmonic Refractive Index Sensors via Synergy between Annealed Nanoparticles and Thin Films. *Nanotechnology* **2020**, *31* (25), 255503. https://doi.org/10.1088/1361-6528/ab7531.

(48) Karakouz, T.; Maoz, B. M.; Lando, G.; Vaskevich, A.; Rubinstein, I. Stabilization of Gold Nanoparticle Films on Glass by Thermal Embedding. *ACS Appl. Mater. Interfaces* **2011**, *3* (4), 978–987. https://doi.org/10.1021/am100878r.

(49) Ryu, M.; Ng, S. H.; Anand, V.; Lundgaard, S.; Hu, J.; Katkus, T.; Appadoo, D.; Vilagosh, Z.; Wood, A. W.; Juodkazis, S.; Morikawa, J. Attenuated Total Reflection at THz Wavelengths: Prospective Use of Total Internal Reflection and Polariscopy. *Applied Sciences* **2021**, *11* (16), 7632. https://doi.org/10.3390/app11167632.

(50) Averett, L. A.; Griffiths, P. R.; Nishikida, K. Effective Path Length in Attenuated Total Reflection Spectroscopy. *Anal. Chem.* **2008**, *80* (8), 3045–3049. https://doi.org/10.1021/ac7025892.

(51) Homola, J.; Yee, S. S.; Gauglitz, G. Surface Plasmon Resonance Sensors: Review. *Sensors and Actuators B: Chemical* **1999**, *54* (1), 3–15. https://doi.org/10.1016/S0925-4005(98)00321-9.

(52) Moirangthem, R. S.; Chang, Y.-C.; Wei, P.-K. Ellipsometry Study on Gold-Nanoparticle-Coated Gold Thin Film for Biosensing Application. *Biomed. Opt. Express, BOE* **2011**, *2* (9), 2569–2576. https://doi.org/10.1364/BOE.2.002569.

(53) Nabok, A.; Tsargorodskaya, A. The Method of Total Internal Reflection Ellipsometry for Thin Film Characterisation and Sensing. *Thin Solid Films* **2008**, *516* (24), 8993–9001. https://doi.org/10.1016/j.tsf.2007.11.077.

(54) Berciaud, S.; Cognet, L.; Tamarat, P.; Lounis, B. Observation of Intrinsic Size Effects in the Optical Response of Individual Gold Nanoparticles. *Nano Lett.* **2005**, *5* (3), 515–518. https://doi.org/10.1021/nl050062t.

(55) Abelès, F. Surface Electromagnetic Waves Ellipsometry. *Surface Science* **1976**, *56*, 237–251. https://doi.org/10.1016/0039-6028(76)90450-7.

(56) Ras, R. H. A.; Schoonheydt, R. A.; Johnston, C. T. Relation between S-Polarized and p-Polarized Internal Reflection Spectra: Application for the Spectral Resolution of Perpendicular Vibrational Modes. *J. Phys. Chem. A* **2007**, *111* (36), 8787–8791. https://doi.org/10.1021/jp073108a.

(57) Wang, H. Plasmonic Refractive Index Sensing Using Strongly Coupled Metal Nanoantennas: Nonlocal Limitations. *Sci Rep* **2018**, *8* (1), 9589. https://doi.org/10.1038/s41598-018-28011-x.

(58) Li, A.; Qiao, X.; Liu, K.; Bai, W.; Wang, T. Hollow Metal Organic Framework Improves the Sensitivity and Anti-Interference of the Detection of Exhaled Volatile Organic Compounds. *Advanced Functional Materials n/a* (n/a), 2202805. https://doi.org/10.1002/adfm.202202805.

(59) He, C.; Liu, L.; Korposh, S.; Correia, R.; Morgan, S. P. Volatile Organic Compound Vapour Measurements Using a Localised Surface Plasmon Resonance Optical Fibre Sensor Decorated with a Metal-Organic Framework. *Sensors* **2021**, *21* (4), 1420. https://doi.org/10.3390/s21041420.

(60) Kreno, L. E.; Hupp, J. T.; Van Duyne, R. P. Metal−Organic Framework Thin Film for Enhanced Localized Surface Plasmon Resonance Gas Sensing. *Anal. Chem.* **2010**, *82* (19), 8042–8046. https://doi.org/10.1021/ac102127p.

(61) Koh, C. S. L.; Lee, H. K.; Han, X.; Sim, H. Y. F.; Ling, X. Y. Plasmonic Nose: Integrating the MOF-Enabled Molecular Preconcentration Effect with a Plasmonic Array for Recognition of Molecular-Level Volatile Organic Compounds. *Chem. Commun.* **2018**, *54*

(20), 2546–2549. https://doi.org/10.1039/C8CC00564H.

(62) Chong, X.; Kim, K.; Zhang, Y.; Li, E.; Ohodnicki, P. R.; Chang, C.-H.; Wang, A. X. Plasmonic Nanopatch Array with Integrated Metal–Organic Framework for Enhanced Infrared Absorption Gas Sensing. *Nanotechnology* **2017**, *28* (26), 26LT01. https://doi.org/10.1088/1361-6528/aa7433.